\documentclass[twocolumn,showpacs,preprintnumbers,amsmath,amssymb]{revtex4}

\usepackage{graphicx}
\usepackage{dcolumn}
\usepackage{bm}

\begin{document}

\preprint{APS/123-QED}
\title{Low temperature specific heat of La$_{3}$Pd$_{4}$Ge$_{4}$ with U$_{3}$Ni$_{4}$Si$_{4}$-type structure}

\author{S.~Kasahara}%
\email{KASAHARA.Shigeru@nims.go.jp} 
\homepage{http://www.nims.jp/filmcrystal/kasahara/kasahara.html}
\author{H.~Fujii}
\author{T.~Mochiku}
\author{H.~Takeya}
\author{K.~Hirata}

\affiliation{
Superconducting Materials Center, National Institute for Materials Science, Sengen 1-2-1, Tsukuba,  Ibaraki 305-0047, Japan
}

\date{\today}

\begin{abstract}
Low temperature specific heat has been investigated in a novel ternary superconductor La$_{3}$Pd$_{4}$Ge$_{4}$ with an U$_{3}$Ni$_{4}$Si$_{4}$-type structure consisting of the alternating BaAl$_{4}$ (ThCr$_{2}$Si$_{2}$)- and AlB$_{2}$-type layers. A comparative study with the related ThCr$_{2}$Si$_{2}$-type superconductor LaPd$_{2}$Ge$_{2}$, one of the layers in La$_{3}$Pd$_{4}$Ge$_{4}$, is also presented. From the normal state specific heat, the Sommerfeld coefficient $\gamma_{n} = 27.0$ mJ/mol K$^2$ and the Debye temperature $\Theta_{\rm D}$ = 256 K are derived for the La$_{3}$Pd$_{4}$Ge$_{4}$, while those for the LaPd$_{2}$Ge$_{2}$ are $\gamma_{n} =8.26$ mJ/mol K$^2$ and $\Theta_{\rm D}$ = 291 K. The La$_{3}$Pd$_{4}$Ge$_{4}$ has moderately high electronic density of state at the Fermi level. Electronic contribution on the specific heat, $C_{\rm el}$, in each compound is well described by the BCS behavior, suggesting that both of the La$_{3}$Pd$_{4}$Ge$_{4}$ and the LaPd$_{2}$Ge$_{2}$ have fully opened isotropic gap in the superconducting state. 
\end{abstract}

\maketitle

\section{Introduction}

Ternary intermetallic compounds with ThCr$_{2}$Si$_{2}$-type structure, a BaAl$_{4}$-type derivative one, have been studied extensively.~\cite{Ban, Just} Numerous materials have been found in this system. Many of these compounds are crystallized with rare-earth ions, including a number of superconducting and magnetic materials. As represented by CeCu$_{2}$Si$_{2}$, known as the first heavy fermion material showing superconductivity, 4f-electrons by the rare-earth ions sometimes bring exotic properties.~\cite{Steglich_CeCu2Si2} What is interesting here is the existence of further derivative structures. CaBe$_{2}$Ge$_{2}$-~\cite{Eisenmann, LaIr2Si2} and BaNiSn$_{3}$-type structured compounds~\cite{DrrscheidtSchifer, CeRhSi3} are good examples. Especially, the latter type compounds, as represented by CeRhSi$_{3}$ and CeIrSi$_{3}$, have attracted recent interests since their structure lacks the mirror plane normal to the $c$-axis, leading to a possible admixture of spin-singlet and triplet pairing.~\cite{CeRhSi3, Bauer_CePt3Si} Among the ThCr$_{2}$Si$_{2}$-based derivatives, quaternary intermetallic borocarbides, $R$Ni$_{2}$B$_{2}$C, show superconductivity at moderately high transition temperatures ($T_{\rm c}$s).~\cite{Cava_YPd2B2C, Cava_LnNi2B2C, Cava_RPt2B2C, Muller} The superconductivity in YPd$_{2}$B$_{2}$C appears at 23 K which is the highest $T_{\rm c}$ in the ThCr$_{2}$Si$_{2}$-type derivatives. The crystal structure of the borocarbides is interstitially filled ThCr$_{2}$Si$_{2}$-type structure by the carbon atoms. 

Alternatively, the superconductivity in the intermetallic compounds has been focused on especially after the discovery of MgB$_{2}$ with the $T_{\rm c}$ as high as 39 K; the highest one over a number of intermetallics.~\cite{Nagamatsu_MgB2} The crystal structure of the MgB$_{2}$ is AlB$_{2}$-type, where the boron atoms form a two-dimensional honeycomb network. 

Recently, novel ternary superconductors La$_3$Pd$_{4}$Ge$_{4}$ and La$_{3}$Pd$_{4}$Si$_{4}$ with $T_{\rm c}$s of 2.75 and 2.15 K have been found.~\cite{Fujii_PRB, Fujii_JPCM} The crystal structure of these compounds is identical to that of U$_{3}$Ni$_{4}$Si$_{4}$, consisting of alternating units of the BaAl$_{4}$ (ThCr$_{2}$Si$_{2}$)- and the AlB$_{2}$-type layers.~\cite{Yarmolyuk_U3Ni4Si4}  Therefore, this type of structure is also considered as one of the further derivatives of the ThCr$_{2}$Si$_{2}$-type one. At present, less than 10 compounds have been found with the U$_{3}$Ni$_{4}$Si$_{4}$-type structure. Among them, only the La$_{3}$Pd$_{4}$Ge$_{4}$ and the La$_{3}$Pd$_{4}$Si$_{4}$ have been found as the superconductors, so far. Contrarily, the $T_{\rm c}$s of the original ThCr$_{2}$Si$_{2}$-type compounds, LaPd$_{2}$Ge$_{2}$ and LaPd$_{2}$Si$_{2}$, are 1.17 and 0.39 K, respectively.~\cite{Hull,Palstra}  Both of the U$_{3}$Ni$_{4}$Si$_{4}$-type compounds, La$_{3}$Pd$_{4}$Ge$_{4}$ and the La$_{3}$Pd$_{4}$Si$_{4}$, show superconductivity at higher $T_{\rm c}$s than the ThCr$_{2}$Si$_{2}$-type ones. The U$_{3}$Ni$_{4}$Si$_{4}$-type system appears to be a new candidate of the ThCr$_{2}$Si$_{2}$-type derivatives for investigating the exotic superconductivity.

In this work, we present a comparative study on the thermodynamic properties of the La$_{3}$Pd$_{4}$Ge$_{4}$ and the LaPd$_{2}$Ge$_{2}$ to shed light on the superconducting natures. The Sommerfeld coefficient $\gamma_{\rm n}$ and the Debye temperature $\Theta_{\rm D}$ are derived from the normal state specific heat. It turned out that the La$_{3}$Pd$_{4}$Ge$_{4}$ has moderately high $\gamma_{\rm n}$ and electronic density of state at the Fermi level, $N(E_{\rm F})$. The gap symmetry of these compounds is also discussed based on the electronic specific heat, $C_{\rm el}$, in the superconducting state. In the superconducting state, the $C_{\rm el}$ in each compound is well described by the exponential behavior. Jumps in the specific heat, $\Delta C$s, at the $T_{\rm c}$s are slightly smaller than the BCS requests, but within the framework. Additionally, specific heat measurements under a magnetic field, $H$, revealed that the Sommerfeld coefficient in the superconducting state,  $\gamma_{\rm s}(H)$, for the La$_{3}$Pd$_{4}$Ge$_{4}$ changes linearly with the increase of $H$. These results suggest that the La$_{3}$Pd$_{4}$Ge$_{4}$ and the LaPd$_{2}$Ge$_{2}$ have fully opened isotropic gap structures in the superconducting states. The increase of $T_{\rm c}$ in the La$_{3}$Pd$_{4}$Ge$_{4}$ and the moderately high $N(E_{\rm F})$ due to the U$_{3}$Ni$_{4}$Si$_{4}$-type structure are also discussed based on the specific heat results. An interesting superconductivity in the intermetallics with the U$_{3}$Ni$_{4}$Si$_{4}$-type structure is focused on.

\section{experimental}

\begin{figure}[b]
\begin{center}\leavevmode
\includegraphics[width=0.8\linewidth]{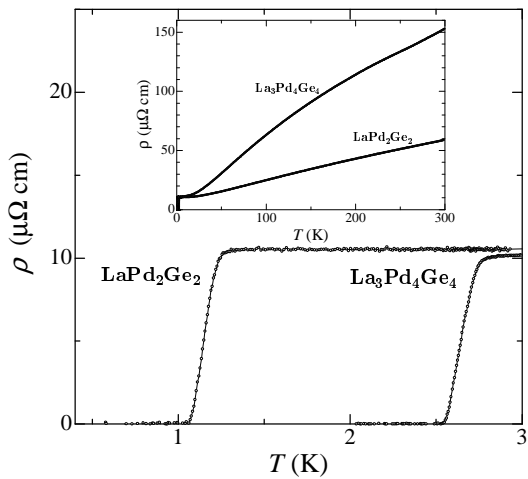}
\caption{
Resistivity curves of the La$_{3}$Pd$_{4}$Ge$_{4}$ and the LaPd$_{2}$Ge$_{2}$ samples at low temperature. 
The La$_{3}$Pd$_{4}$Ge$_{4}$ and the LaPd$_{2}$Ge$_{2}$ samples show zero resisitivity at 2.5 and 1.1 K, respectively. 
The inset shows the resistivity curves up to 300 K. 
}
\label{RT}
\end{center}
\end{figure}

Samples of the La$_{3}$Pd$_{4}$Ge$_{4}$ and the LaPd$_{2}$Ge$_{2}$ were synthesized by the standard arc-melting technique with the stoichiometric ratio of pure elements, La(3N), Pd(3N) and Ge(3N), under Ar gas atmosphere. The arc-melted buttons were annealed at 900-1000$^\circ$C in vacuum, typically for 1 week. X-ray diffraction patterns in the La$_{3}$Pd$_{4}$Ge$_{4}$ clarified that the crystal structure is identical to the U$_{3}$Ni$_{4}$Si$_{4}$-type, with a few percent impurities mainly attributed to the LaPd$_{2}$Ge$_{2}$. Contrarily, as for the LaPd$_{2}$Ge$_{2}$, a single phased sample is obtained. 
Resistvity measurements in the La$_{3}$Pd$_{4}$Ge$_{4}$ and the LaPd$_{2}$Ge$_{2}$ were carried out by the standard 4-probes technique (Fig.~\ref{RT}).
The residual resistivity of the present samples is about 10 $\mu\Omega\cdot$cm, which is comparable to the previous reports.~\cite{Fujii_PRB,Hull} 
Specific heat measurements with thermal relaxation method were performed in a temperature range from 0.4 to 5.0 K on platelet-shaped polycrystals cut from the annealed buttons. 
Clear specific heat jumps are observed at the onset of 2.5 K for the La$_{3}$Pd$_{4}$Ge$_{4}$, and 1.1 K for the LaPd$_{2}$Ge$_{2}$ samples, which are equivalent to the zero resistivity temperature in the both compounds.

\section{Results and discussion}

\begin{figure}[t]
\begin{center}\leavevmode
\includegraphics[width=0.7\linewidth]{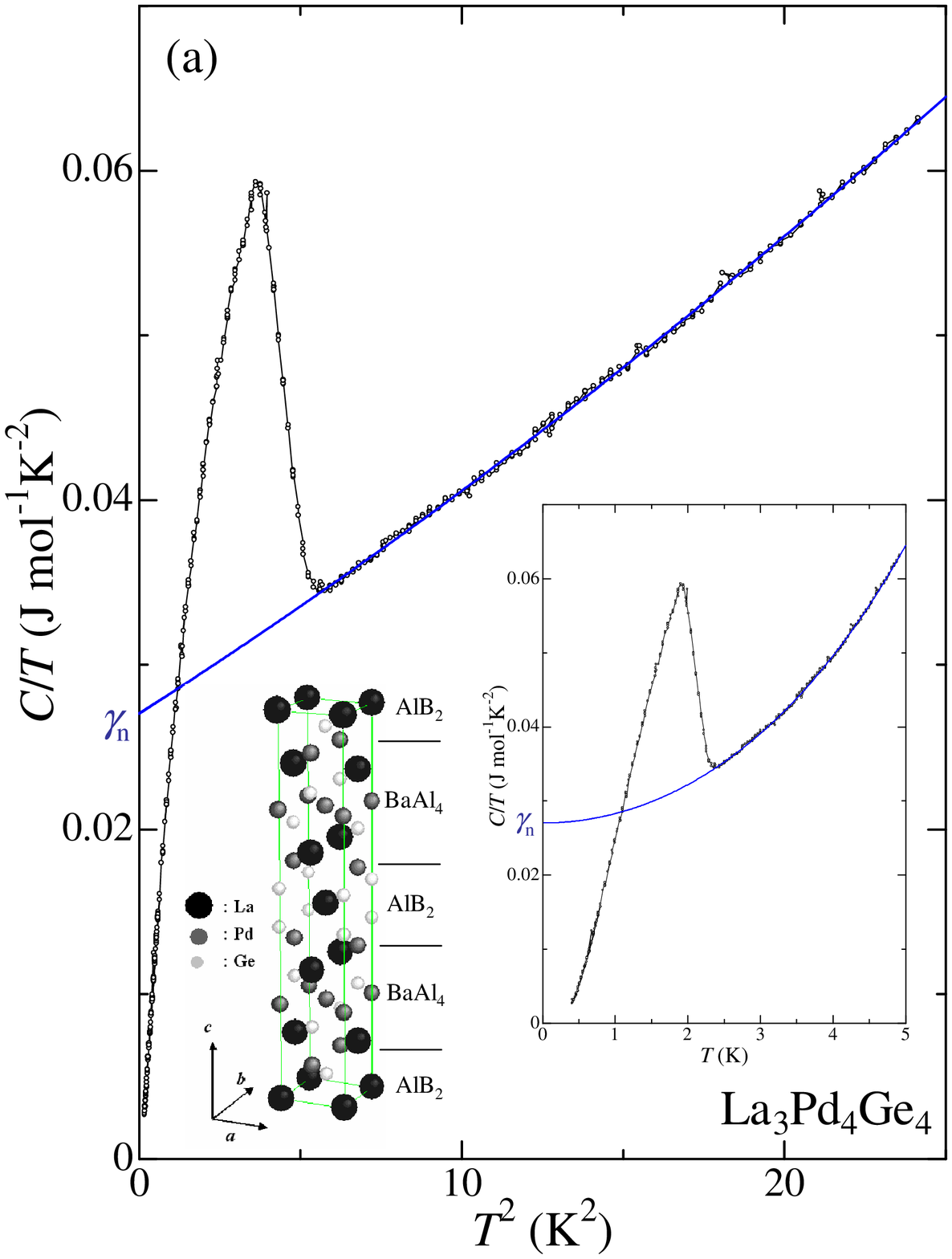}

\vspace{10pt}
\includegraphics[width=0.7\linewidth]{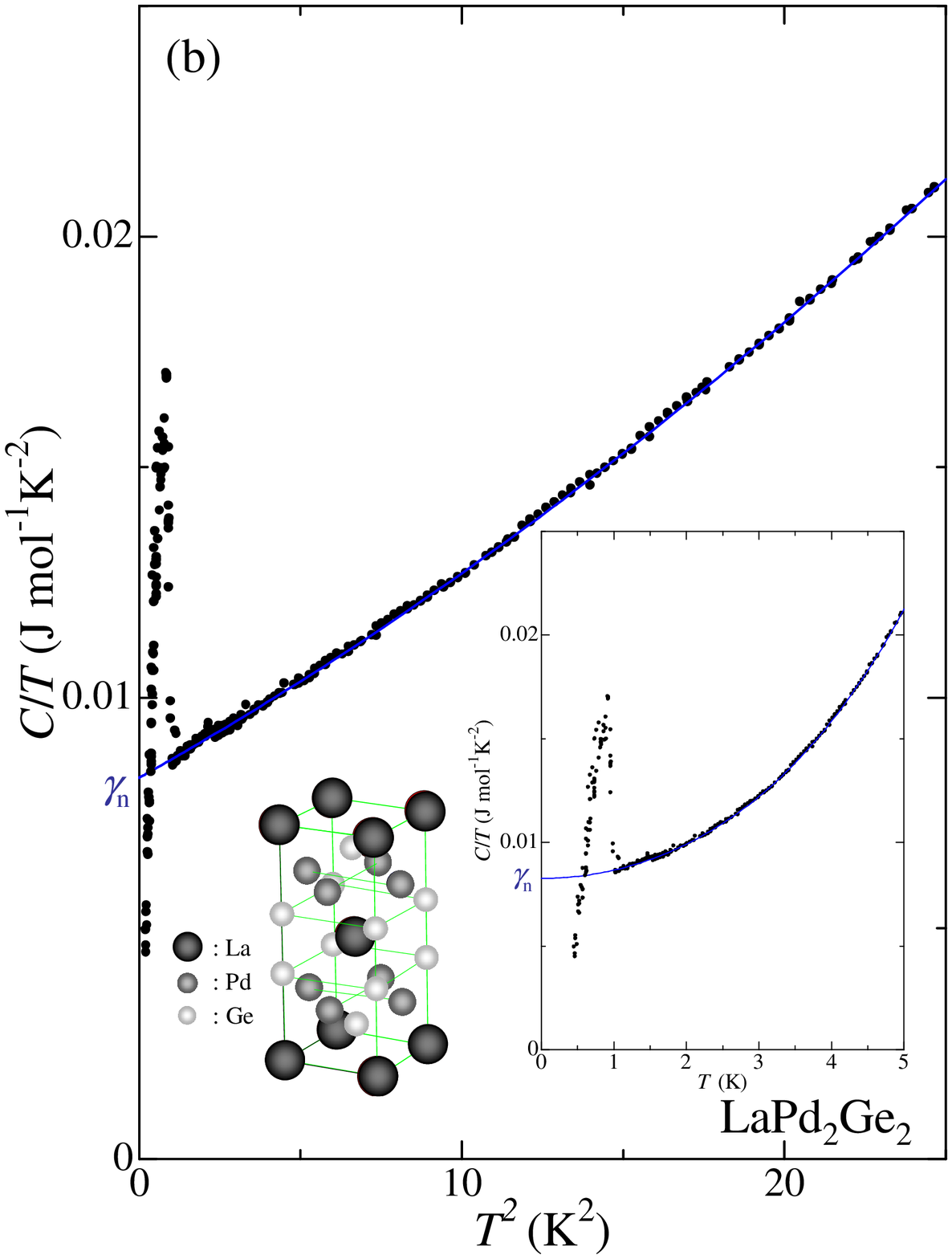}
\caption{
Low temperature specific heat in the La$_{3}$Pd$_{4}$Ge$_{4}$ (a), and the LaPd$_{2}$Ge$_{2}$ (b). 
The main panels display $C/T$ versus $T^{2}$ plot, while the right insets show $C/T$ versus $T$. 
The left insets are the crystal structure of the La$_{3}$Pd$_{4}$Ge$_{4}$ and the LaPd$_{2}$Ge$_{2}$. 
The solid lines describe the fits to the data by $C = \gamma_{n}T + \beta T^{3} + \delta T^{5}$.
The parameters $\gamma_{n}$, $\beta$, $\delta$ in both superconductors are summarized in Tab.~\ref{table1}. 
}
\label{C}
\end{center}
\end{figure}

Figure~\ref{C}(a) shows the temperature ($T$) divided specific heat ($C$) of the La$_{3}$Pd$_{4}$Ge$_{4}$ in $C/T$ versus $T^{2}$, and also $C/T$ versus $T$ plots (the right inset). The left inset shows the crystal structure of the La$_{3}$Pd$_{4}$Ge$_{4}$. A clear specific heat jump in the present sample appears at 2.5 K, which is almost equal to the zero resistance temperature in the resistivity measurements in the Fig.~\ref{RT}. The jump suggests a bulk nature of the superconductivity. 
No other anomaly due to a few percent of the LaPd$_{2}$Ge$_{2}$ as the secondary impurity phase was observed. The normal state specific heat in the La$_{3}$Pd$_{4}$Ge$_{4}$ is well fitted by the linear combination of the electronic contribution $C_{\rm el} = \gamma_{\rm n}T$ and the phonon one $C_{\rm ph} = \beta T^{3} + \delta T^{5}$, with $\gamma_{n} =$ 27.0 mJ/mol K$^2$, $\beta$ = 1.27 mJ/mol K$^{4}$, $\delta =$ 12.7 mJ/mol K$^{6}$, as described by the solid line. The Debye temperature $\Theta_{\rm D} = 256$ K is derived from the relationship, $\beta = (12\pi^{4}/5)Nk_{\rm B}(1/\Theta_{\rm D})^{3}$. The electron-phonon coupling constant $\lambda_{\rm ph}$ is given by the McMillan's formula 
\begin{equation}
T_{c} = \frac{\Theta_{\rm D}}{1.45}\exp\left[-\frac{1.04(1+\lambda_{\rm ph})}{\lambda_{\rm ph}-\mu^{\ast}(1+0.62\lambda_{\rm ph})} \right]
\end{equation}
where $\mu^{\ast}$ is the Coulomb pseudopotential~\cite{McMillan}. Taking $\mu^{\ast} = $ 0.10 - 0.13, we obtain $\lambda_{\rm ph} =$ 0.48 - 0.55, suggesting that the La$_{3}$Pd$_{4}$Ge$_{4}$ is a weak-coupled superconductor. 

A comparative study on the LaPd$_{2}$Ge$_{2}$ following the same procedure as the La$_{3}$Pd$_{4}$Ge$_{4}$ is described in Fig.~\ref{C}(b). $\gamma_{\rm n}$, $\beta$, and $\delta$ are estimated as 8.26 mJ/mol~K$^{2}$, 0.393 mJ/mol~K$^{4}$, 5.06 $\mu$J/mol K$^{6}$, respectively. The value of $\gamma_{\rm n}$ for the La$_{3}$Pd$_{4}$Ge$_{4}$ is more than three times larger than that for the LaPd$_{2}$Ge$_{2}$. On the other hand, $\Theta_{\rm D}$ for the LaPd$_{2}$Ge$_{2}$ is derived as 291 K, which is higher than that for the La$_{3}$Pd$_{4}$Ge$_{4}$. The electron-phonon coupling constant in the LaPd$_{2}$Ge$_{2}$ is estimated as $\lambda_{\rm ph} = 0.39-0.45$ for $\mu^{\ast} = 0.10-0.13$, smaller than that in the La$_{3}$Pd$_{4}$Ge$_{4}$.

The electronic density of states at the Fermi level, $N(E_{\rm F})$, is given by the following relationship 
\begin{equation}
\gamma_{\rm n} = \frac{2}{3}\pi^{2}k_{\rm B}^{2}(1 + \lambda_{\rm ph})N(E_{\rm F}). 
\end{equation}
Evidently, $N(E_{\rm F})$ can be derived from the given $\gamma_{\rm n}$ and the $\lambda_{\rm ph}$. Taking $\gamma_{\rm n} = 27.0$ mJ/mol K$^{2}$ and $\lambda_{\rm ph} = 0.48$-0.55 for the La$_{3}$Pd$_{4}$Ge$_{4}$, $N(E_{\rm F}) =$ 3.86 - 3.69 states/(eV-formula unit). Contrarily, $N(E_{\rm F})$ in the LaPd$_{2}$Ge$_{2}$ is derived as 1.69 - 1.63 states/(eV-formula unit) for the value $\gamma_{\rm n} = 8.26$ mJ/mol K$^{2}$ and $\lambda_{\rm ph} = 0.39$-0.45. The results suggest that, for the unit cell, the $N(E_{\rm F})$ in the U$_{3}$Ni$_{4}$Si$_{4}$-type structured La$_{3}$Pd$_{4}$Ge$_{4}$ is more than three times higher than that in the ThCr$_{2}$Si$_{2}$-type compounds, LaPd$_{2}$Ge$_{2}$. 
Assuming even contributions on the $N(E_{\rm F})$ by all the atoms in the unit cell, electronic density of states per atom, $n(E_{\rm F})$, for the La$_{3}$Pd$_{4}$Ge$_{4}$ is calculated as 0.351-0.333 states/(eV-atom), while that for the LaPd$_{2}$Ge$_{2}$ is 0.338-0.326 states/(eV-atom). 
The La$_{3}$Pd$_{4}$Ge$_{4}$ has a few percent higher $n(E_{\rm F})$ than the LaPd$_{2}$Ge$_{2}$.

\begin{figure}[t]
\begin{center}\leavevmode
\includegraphics[width=0.70\linewidth]{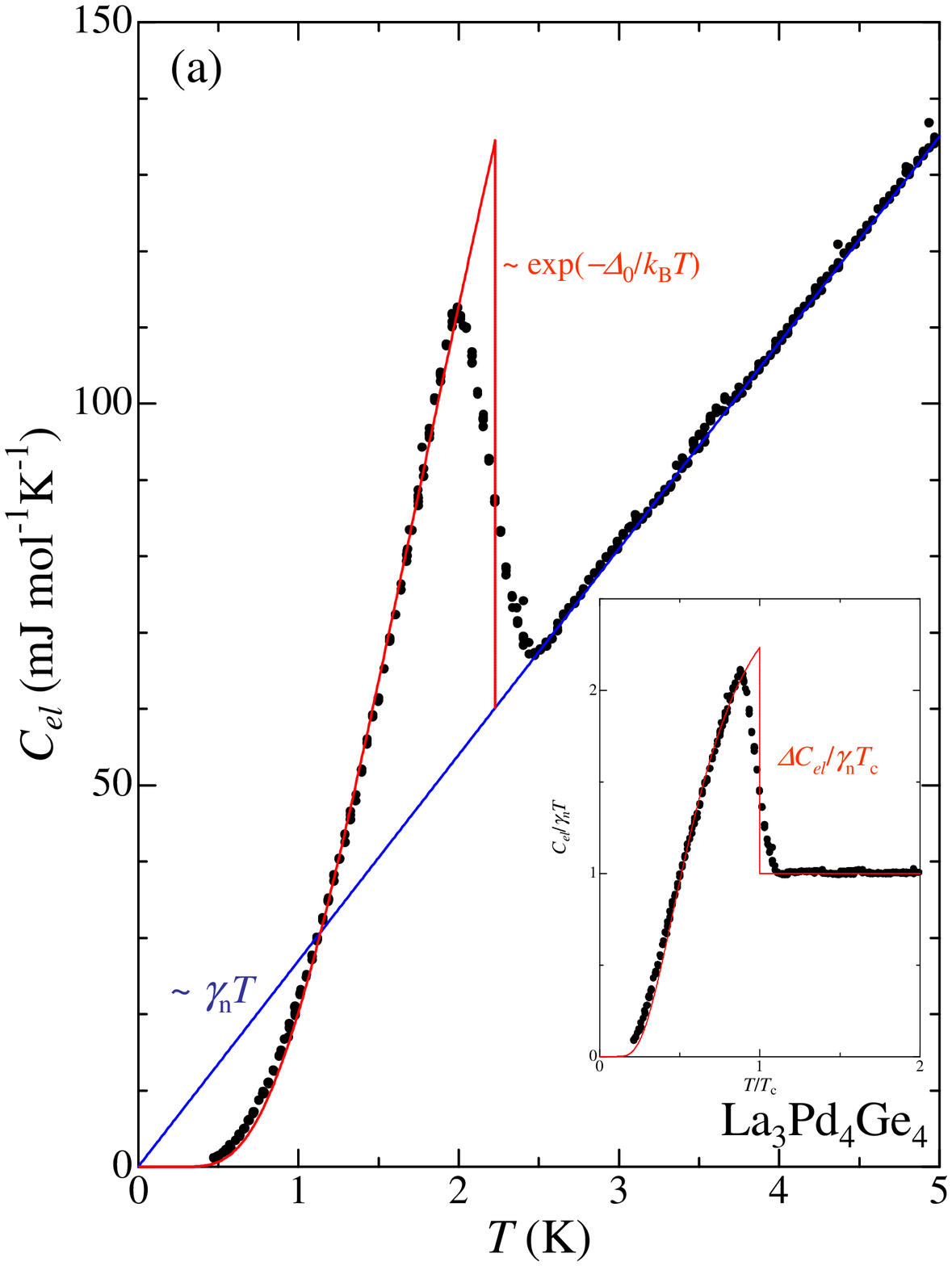}

\vspace{10pt}
\includegraphics[width=0.70\linewidth]{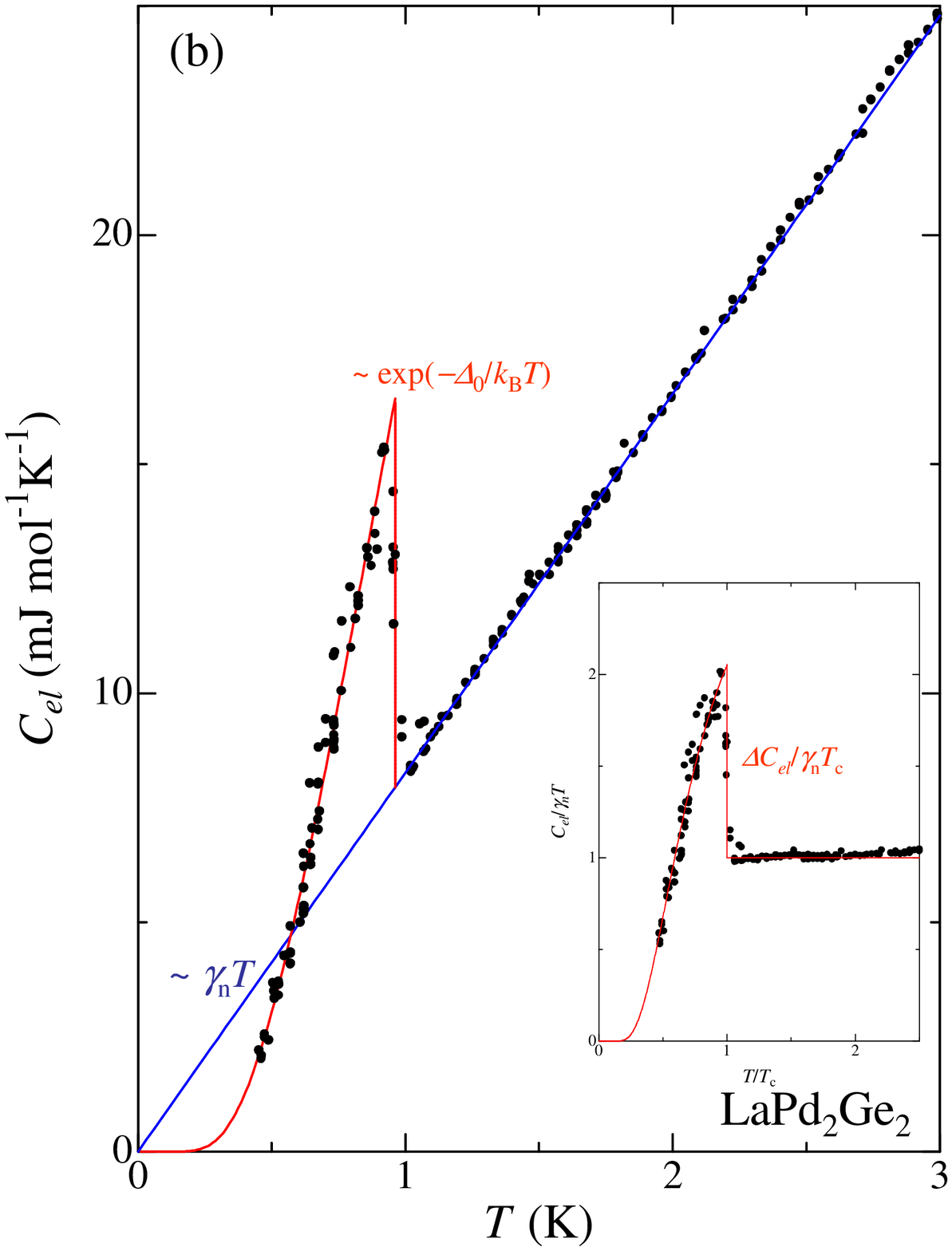}

\caption{
The electronic specific heat, $C_{\rm el}$, in La$_{3}$Pd$_{4}$Ge$_{4}$ (main panel), and the normalized plot $C_{\rm el}/\gamma_{n}T$ (the inset). 
$C_{\rm el}$ in the superconducting state shows good agreement with the exponential behavior, consistent with the gap excitation of the quasi-particles described by the BCS framework.  
}
\label{Cel}
\end{center}
\end{figure}

\begin{table*}[t]
\begin{center}
\caption{Normal and the superconducting parameters of the La$_{3}$Pd$_{4}$Ge$_{4}$ and LaPd$_{2}$Ge$_{2}$ derived from the specific heat results.}
\begin{tabular}{l c c c c c c c }
\hline
& & $T_{\rm c}^{\rm on}$ (K) & $T_{\rm c}^{\rm th}$ (K) & $\gamma_{n}$ (mJ/mol K$^{2}$) & $\beta$ (mJ/mol K$^{4}$) & $\delta$ ($\mu$J/mol K$^{6}$)  & $\Theta_{\rm D}$ (K) \\
\hline
\hline
La$_{3}$Pd$_{4}$Ge$_{4}$ & & 2.50 & 2.23 & 27.0 & 1.27 & 12.7 & 256 \\
\hline
LaPd$_{2}$Ge$_{2}$ & & 1.10 & 0.96 & 8.26 & 0.393 & 5.05  & 291 \\
\hline\\
\\
\hline
&& $\lambda_{\rm ph}$ & $N(E_{F})$ (states/eV~f.~u.~) & $2\Delta^{\rm exp}_{0}/k_{\rm B}T_{\rm c}$ & $\Delta^{\rm exp}_{0}$ & $\Delta^{\rm BCS}_{0}$ & $\Delta C/\gamma_{n}T_{c}^{\rm th}$ \\
\hline
\hline
La$_{3}$Pd$_{4}$Ge$_{4}$ && 0.48-0.55 & 3.86-3.69 & 3.08 & 0.30 & 0.34 & 1.24 \\
\hline
LaPd$_{2}$Ge$_{2}$  && 0.39-0.45 & 1.69-1.63 & 3.65 & 0.15 & 0.15 & 1.05 \\
\hline
\end{tabular}
\label{table1}
\end{center}
\end{table*}

\begin{figure*}[t]
\begin{center}\leavevmode
\includegraphics[width=0.5\linewidth]{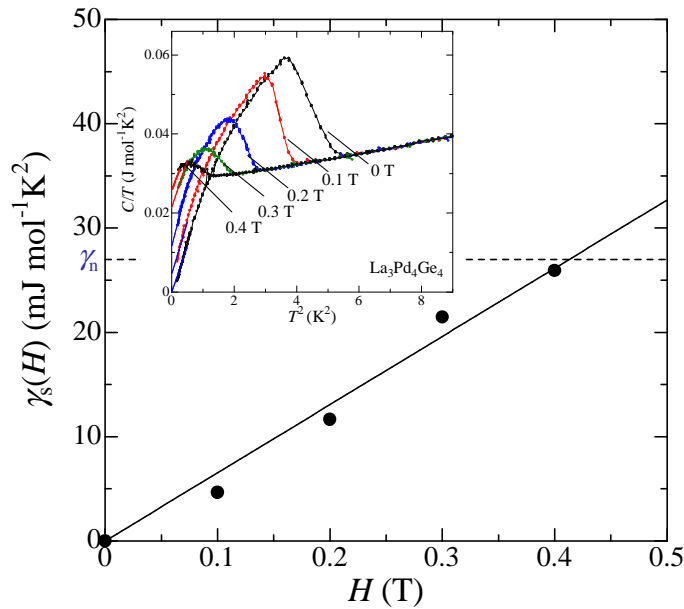}
\caption{
Magnetic field dependence of the Sommerfeld coefficient in the superconducting state, $\gamma_{s}(H)$, defined by extrapolating the $C/T$ vs $T^{2}$ plot (the inset) below $T^{2} \leq 0.5$ K$^{2}$. The solid line shows a function $\gamma_{s}(H) = \gamma_{s}(0) + \gamma_{n}H/H_{\rm c2}$. The field dependence of $\gamma_{s}(H)$ shows good agreement with the conventional BCS superconductor with an isotropic gap. 
}
\label{gamma_s}
\end{center}
\end{figure*}

The electronic contribution of the specific heat, $C_{\rm el} = C - \beta T^{3} -\delta T^{5}$, in the La$_{3}$Pd$_{4}$Ge$_{4}$ is described in Fig.\ref{Cel}(a). The main panel and the inset show $C_{\rm el}$ versus $T$, and $C_{\rm el}/\gamma_{\rm n}T$ versus $T/T_{\rm c}^{\rm th}$ for the La$_{3}$Pd$_{4}$Ge$_{4}$, respectively. Here, $T_{\rm c}^{\rm th}$ denotes the thermodynamic critical temperature for the superconductivity.~\cite{note}  The $C_{\rm el}$ in the superconducting state shows a good agreement with a thermally excited exponential behavior, $C_{\rm el} \sim \exp(-\Delta_{0}/k_{\rm B}T)$, as expected for the BCS superconductivity. The superconducting gap determined by the experiments is $\Delta_{0}^{\rm exp} =$ 0.30 meV, which is slightly smaller than the BCS request, $\Delta_{0}^{\rm BCS}$ = 0.33 meV, expected for the $T_{\rm c}^{\rm th} \sim $ 2.23 K.~\cite{note} The gap $\Delta_{0}^{\rm exp}$ gives a value $2\Delta_{0}^{\rm exp}/k_{\rm B}T_{\rm c}^{\rm th} \sim 3.08$ and the electronic specific heat jump at the superconducting transition brings $\Delta C_{\rm el}/\gamma_{\rm n}T_{\rm c}^{\rm th} \sim 1.24$. Although these values are smaller than the BCS requests of $2\Delta_{0}^{\rm exp}/k_{\rm B}T_{\rm c}^{\rm th}$ = 3.53 and $\Delta C_{\rm el}/\gamma_{\rm n}T_{\rm c}^{\rm th} = 1.43$, the results suggest that the La$_{3}$Pd$_{4}$Ge$_{4}$ has fully opened isotropic gap in the superconducting order parameter with phonon-mediated pairing mechanism. 

In the same way, $C_{\rm el}$ for the LaPd$_{2}$Ge$_{2}$ is analyzed [Fig.~\ref{Cel}(b)]. The values of $\Delta C_{\rm el}/\gamma_{\rm n}T_{\rm c}^{\rm th} \sim 1.05$ and $\Delta_{\rm 0}^{\rm exp} \sim 0.15$ meV are derived, also consistent with the BCS request of $\Delta_{\rm 0}^{\rm BCS}$ = 0.15 meV. The value $2\Delta_{0}^{\rm exp}/k_{\rm B}T_{\rm c}^{\rm th}$ is given as 3.65. The parameters derived from the specific results are summarized in Tab.~\ref{table1}

In the superconducting state, magnetic field dependence of the Sommerfeld coefficient, $\gamma_{\rm s}(H)$, also provides an information on the gap symmetry. Figure~\ref{gamma_s} describes field evolution of the low temperature specific heat in the La$_{3}$Pd$_{4}$Ge$_{4}$ (the inset) and the $\gamma_{\rm s}(H)$ (main panel). The $\gamma_{\rm s}(H)$ is determined by extrapolating the $C/T$ vs $T^{2}$ curves at $T^{2} \leq 0.5$ K$^2$. Since the LaPd$_{2}$Ge$_{2}$ is a type-I superconductor, the structure in the specific heat is rapidly suppressed by the application of the magnetic field (not shown). The field dependence of the $\gamma_{s}(H)$ in the La$_{3}$Pd$_{4}$Ge$_{4}$ is well described as $\gamma_{s}(H) = \gamma_{s}(0) + \gamma_{n}H/H_{\rm c2}$, which is in marked contrast to the behavior $\gamma_{s}(H) = \gamma_{s}(0) + \gamma_{n}(H/H_{\rm c2})^{1/2}$, expected for nodal superconductors.~\cite{Volovik} As well as the exponential behavior in the $C_{\rm el}$, the results again show that the La$_{3}$Pd$_{4}$Ge$_{4}$ is a fully gapped superconductor, in which $\gamma_{s}(H)$ is proportional to the density of vortices resulting in $\gamma_{s}(H) \propto H$.

Let us compare the superconducting natures in the La$_{3}$Pd$_{4}$Ge$_{4}$ and the LaPd$_{2}$Ge$_{2}$, especially the enhancement of the  $T_{\rm c}$ in the La$_{3}$Pd$_{4}$Ge$_{4}$ by taking the U$_{3}$Ni$_{4}$Si$_{4}$-type structure. The $\Theta_{\rm D}$ for the La$_{3}$Pd$_{4}$Ge$_{4}$ is about 35 K lower than that for the LaPd$_{2}$Ge$_{2}$, while $T_{\rm c}$ for the La$_{3}$Pd$_{4}$Ge$_{4}$ is more than two times higher than that for the LaPd$_{2}$Ge$_{2}$. This is attributed to the difference in the electron-phonon coupling as the La$_{3}$Pd$_{4}$Ge$_{4}$ has about 20\% larger $\lambda_{\rm ph}$ than the LaPd$_{2}$Ge$_{2}$. 
Contrarily, the $\gamma_{\rm n}$ in the La$_{3}$Pd$_{4}$Ge$_{4}$ is more than three times higher than that in the LaPd$_{2}$Ge$_{2}$. 
The moderately high $\gamma_{\rm n}$ in the La$_{3}$Pd$_{4}$Ge$_{4}$ leads to the high $N(E_{\rm F})$. 
The structural change from the ThCr$_{2}$Si$_{2}$-type structure, and the insertion of the AlB$_{2}$-type layer can cause an extra contribution on the Fermi level. 
Such electrons at the Fermi level may play an important role on the higher $T_{\rm c}$ of the superconductivity. 

The existence of the AlB$_{2}$-type layers in the U$_{3}$Ni$_{4}$Ge$_{4}$-type structure may be a key for the superconductivity. So far, it has been reported that most of the rare-earths substitutions to the La fail to stabilize the U$_{3}$Ni$_{4}$Si$_{4}$-type structure. For instance, $Ln_{3}$Pd$_{4}$Ge$_{4}$ ($Ln =$ Y, Gd, Tb, Dy, Ho, Er, Tm and Yb) take so called Gd$_{3}$Cu$_{4}$Ge$_{4}$-type structure,~\cite{Gd3Cu4Ge4_1,Gd3Cu4Ge4_2,Gd3Cu4Ge4_3,Gd3Cu4Ge4_4,Gd3Cu4Ge4_5,Gd3Cu4Ge4_6} in which cages of Pd-Ge are stabilized instead of the Pd-Ge networks in La$_{3}$Pd$_{4}$Ge$_{4}$. No AlB$_{2}$-type layer is realized in these compounds, and interestingly, no superconductivity appears there. This fact implies that the existence of the AlB$_{2}$-type layers, especially the three-fold coordination of the Pd to the Ge and the 2-dimensional network of the Pd-Ge may have some relations to the superconductivity.~\cite{Mochiku} 

In addition to the present results, we have very recently found novel superconductivity in La$_{3}$Ni$_{4}$Si$_{4}$ and La$_{3}$Ni$_{4}$Ge$_{4}$ at 1.0 and 0.7 K, which are also the U$_{3}$Ni$_{4}$Si$_{4}$-type intermetallics, even though the ThCr$_{2}$Si$_{2}$-type original compounds LaNi$_{2}$Si$_{2}$ and LaNi$_{2}$Ge$_{2}$ do not show superconductivity at least down to 300 mK.~\cite{Fujii_LaNiX} Although other known U$_{3}$Ni$_{4}$Si$_{4}$-type compounds, La$_{3}$Rh$_{4}$Ge$_{4}$, Ce$_{3}$Rh$_{4}$Ge$_{4}$, Ce$_{3}$Ph$_{3}$IrGe$_{4}$, and U$_{3}$Ni$_{4}$Si$_{4}$ are still uninvestigated, there possibly underlies similar changes in the electronic states. The present experiments on the specific heat revealed that both of the La$_{3}$Pd$_{4}$Ge$_{4}$ and the LaPd$_{2}$Ge$_{2}$ show the superconductivity with isotropic gap structures. The results seem fairly reasonable since the lanthanum does not have 4f-electrons. If the lanthanum in the La$_{3}$Pd$_{4}$Ge$_{4}$ can be substituted to the other rare-earths, it might bring some exotic properties in the superconducting nature. Further investigations on the U$_{3}$Ni$_{4}$Si$_{4}$-type intermetallics, as well as the band structure calculations, are awaited.

\section{Summary}
In summary, we have investigated the low temperature specific heat in the ternary germanide superconductor La$_{3}$Pd$_{4}$Ge$_{4}$ with the U$_{3}$Ni$_{4}$Si$_{4}$-type structure, and also in the LaPd$_{2}$Ge$_{2}$ with the ThCr$_{2}$Si$_{2}$-type structure. The normal state specific heat revealed that the Sommerfeld coefficient and the Debye temperature in the La$_{3}$Pd$_{4}$Ge$_{4}$ are $\gamma_{\rm n} = 27.0$ mJ/mol~K$^{2}$ and $\Theta_{\rm D} = 256$ K, while those in the LaPd$_{2}$Ge$_{2}$ are $\gamma_{\rm n} = 8.26$ mJ/mol~K$^{2}$ and $\Theta_{\rm D} = 291$ K. The results yield moderately high electronic density of state at the Fermi level, $N(E_{\rm F})$, in the La$_{3}$Pd$_{4}$Ge$_{4}$. In the superconducting state, the electronic specific heat in both compounds show thermally excited exponential behavior with gap $\Delta_{0}^{\rm exp} = 0.30$ meV (La$_{3}$Pd$_{4}$Si$_{4}$) and $\Delta_{0}^{\rm exp}=0.15$ meV (LaPd$_{2}$Ge$_{2}$). The Sommerfeld coefficient in the superconducting state, $\gamma_{\rm s}$, in the La$_{3}$Pd$_{4}$Ge$_{4}$ shows linear field dependence. Those results suggest both of the La$_{3}$Pd$_{4}$Ge$_{4}$ and the LaPd$_{2}$Ge$_{2}$ are the phonon mediated BCS superconductors with isotropic gaps.

\begin{acknowledgments}
This work is partially supported by a Grant-in-Aid Scientific Research from the Ministry of Education, Culture, Sports, Science and Technology (MEXT), Japan. 
\end{acknowledgments}


\begin{thebibliography}{99}
\bibitem{note} The critical temperature, $T_{c}^{th}$, for thermodynamic measurements is defined as a point where the transition structure in the specific heat is equally separated. Therefore, the critical temperature here is different from that in Ref.~\cite{Fujii_PRB} which reports the discovery of superconductivity in La$_{3}$Pd$_{4}$Ge$_{4}$, defining $T_{c}$ as an onset of the diamagnetism.
\end{thebibliography}
\end{document}